\documentclass[doublecol]{epl2}
\title{Electronic reconstruction and enhanced superconductivity at La$_{1.6-x}$Nd$_{0.4}$Sr$_{x}$CuO$_{4}$/La$_{1.55}$Sr$_{0.45}$CuO$_{4}$ bilayer interface}
\shorttitle{Electronic reconstruction and enhanced superconductivity...}

\author{P. K. Rout\inst{1} \and P. C. Joshi\inst{1} \and Rajni Porwal\inst{1,2} \and R. C. Budhani\inst{1,2}\thanks{E-mail:\email{rcb@iitk.ac.in,rcb@nplindia.org}}}
\shortauthor{P. K. Rout \etal}

\institute{\inst{1} Condensed Matter - Low Dimensional Systems Laboratory, Department of Physics, Indian Institute of Technology Kanpur, Kanpur - 208016, India\\
  \inst{2} National Physical Laboratory, Council of Scientific and Industrial Research, New Delhi - 110012, India}

\abstract{We report enhanced superconductivity in bilayer thin films consisting of superconducting La$_{1.6-x}$Nd$_{0.4}$Sr$_{x}$CuO$_{4}$ with 0.06 $\leq x<$ 0.20 and metallic but non-superconducting La$_{1.55}$Sr$_{0.45}$CuO$_{4}$. These bilayers show a maximum increase in superconducting transition temperature ($T_c$) of more than 200 $\%$ for $x$ = 0.06 while no change in $T_c$ is observed for the bilayers with $x\geq$ 0.20. The analysis of the critical current and kinetic inductance data suggests 2-3 unit cells thick interfacial layer electronically perturbed to have a higher $T_c$. A simple charge transfer model with cation intermixing explains the observed $T_c$ in bilayers. Still the unusually large thickness of interfacial superconducting layers can not be explained in terms of this model. We believe the stripe relaxation as well as the proximity effect also influence the superconductivity of the interface.}

\begin{document}
\maketitle

\section{Introduction}
A number of recent experiments and theories have provided a new insight of remarkably different electronic properties of the interfaces between two correlated electron systems as compared to those of the parent compounds in their bulk form \cite{Zubko}. One of the unusual realizations of these effects is the superconductivity (SC) seen at the interface of a certain class of oxides \cite{Gozar, Smadici, Logenov, Yuli, Rout}. A recent experiment by Gozar $et$ $al.$ shows the existence of superconductivity at the interface of a bilayer comprising of La$_{2}$CuO$_{4}$ and La$_{1.55}$Sr$_{0.45}$CuO$_{4}$ films, with neither of them being a superconductor on its own \cite{Gozar}. Moreover, an enhancement of the superconducting transition temperature ($T_c$) has been reported for La$_{2-x}$Sr$_{x}$CuO$_{4}$ and La$_{1.875}$Ba$_{0.125}$CuO$_{4}$ thin films when capped by an overdoped metallic La$_{1.65}$Sr$_{0.35}$CuO$_{4}$ \cite{Yuli}. Similar $T_c$ enhancement in electron doped cuprates has been observed for bilayers and multilayers consisting of underdoped (UD) and overdoped (OD) components \cite{Jin}. The material and process factors responsible for the enhanced $T_c$ have been identified as the substrate induced strain due to lattice mismatch \cite{Locquet}, the presence of excess oxygen \cite{Bozovic}, and the inter-diffusion of ions at the interface \cite{Logenov}. From a theoretical point of view, these results have been interpreted as a consequence of strong pairing interaction of UD layer on the large phase stiffness present in the OD layer, which results in the formation of a more robust superconducting phase in the OD layer \cite{Berg}. Also the theoretical approach which considers the delocalization of carriers at the interfaces due to the electrostatic potentials also seems to be in agreement with many experimental results \cite{Loktev}.

\begin{figure*}[htb]
\begin{center}
%\vskip -1.5cm
%\abovecaptionskip -10cm
\includegraphics [width=17cm]{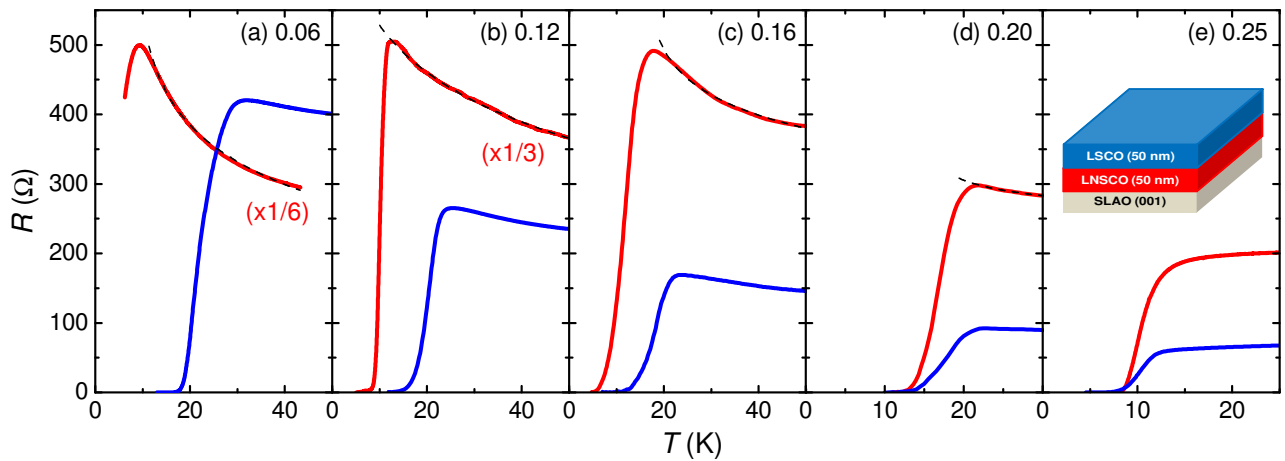}%
\end{center}
\caption{\label{fig1} (Color online) (a-e) The temperature dependent in-plane resistance $R (T)$ for LNSCO (thin red line) and LNSCO/LSCO bilayer (thick blue line) films. The data for LNSCO film in (a) and (b) are multiplied by 1/6 and 1/3 respectively. The black dashed lines show the fits to low temperature resistance upturns in LNSCO films using the relation: $R \propto$ log (1/$T$). The inset of (e) shows the sketch of the LSCO (50 nm)/LNSCO (50 nm)/SLAO bilayer.}
\end{figure*}

The Nd doping in La$_{2-x}$Sr$_{x}$CuO$_{4}$ introduces static charge stripe order and associated spin stripes, which appear to coexist with SC at very low temperatures \cite{Tranquada}. However, in thin epitaxial films of these compounds, the lattice mismatch induced strain can strengthen or weaken the stripe order \cite{Rout, Liu, Tsukada}. In the case of films deposited on SrLaAlO$_4$ (SLAO), a weakening of the stripe pinning potential and thus enhanced stripe fluctuations leading to a $T_c$ larger than the value for bulk samples has been reported \cite{Liu,Rout}. The overdoped La$_{2-x}$Sr$_{x}$CuO$_{4}$ is a conventional Fermi liquid in which the dynamic spin susceptibility scales with $T_c$ and eventually disappears at large $x$ where it ceases to be a superconductor \cite{Wakimoto}. Furthermore it is believed that the metallic nature can originate due to the transverse fluctuation of the charged stripes in such compounds.\cite{Kivelson,Ando} We have fabricated the bilayer thin films of superconducting La$_{1.6-x}$Nd$_{0.4}$Sr$_{x}$CuO$_4$ (LNSCO) and metallic but non-superconducting La$_{1.55}$Sr$_{0.45}$CuO$_4$ (LSCO). This system allows us to investigate how different pairing scales present in these two compounds as well as the potential imbalance across the interface modify the electronic properties of the interface. Moreover, it is still unknown how the overdoped material will influence the stripe order present in Nd doped compound when an interface of these two materials is created. Our study shows a large enhancement of $T_c$ (up to 18.5 K) in the bilayers from its value in case of the bare Nd-doped films. Our analysis of the data strongly indicates that the enhanced $T_c$ is due to the presence of an interfacial superconducting layer of few unit cells thickness in the bilayer.

\section{Experimental Details}
The epitaxial thin films of LSCO and LNSCO along with  LSCO/LNSCO bilayer as sketched in Fig. 1(e) were fabricated on (001) oriented SLAO substrates by pulsed laser deposition at 800 $^{\circ}\mathrm{C}$ in 230 mTorr oxygen pressure. The samples were cooled in atmospheric pressure of oxygen to room temperature with one hour annealing at 500 $^{\circ}\mathrm{C}$ to realize full oxygenation of the structure. The single layer films and each component of the bilayers were 50 nm thick. The present study includes a series of LNSCO and bilayer films with $x$ = 0.06, 0.12, 0.16, 0.20, and 0.25. The crystallographic structure and interface quality of the films were characterized in detail using X-ray diffraction (XRD) and atomic force microscopy (AFM). The $\theta$-2$\theta$ XRD scans clearly revealed the presence of both LSCO and LNSCO layers in the bilayer. In combination with $\omega$ and $\phi$ scans, we have established a high quality epitaxial growth of these structures with $c$-axis along the normal to the plane of the substrate. The low-angle oscillations seen in X-ray reflectivity curves are used to determine the thickness of the films which are consistent with the nominal film thickness as determined from the growth rate. The AFM scans over an area of 2$\times$2 $\mu m^2$ showed a typical root-mean-square roughness of 1.3 nm, which is about one unit cell thick ($c \approx$ 1.3 nm). The fitting to reflectivity curve using a genetic algorithm yields the interface roughness of (1.4 $\pm$ 0.1) nm and the surface roughness of (1.7 $\pm$ 0.1) nm \cite{Dane}. The in-plane transport measurements were performed in four-probe geometry. The ac screening measurements were carried out with a two-coil mutual inductance method \cite{Jeanneret}.

\begin{figure}[tb]
\begin{center}
%\vskip -1.5cm
%\abovecaptionskip -10cm
\includegraphics [width=8cm]{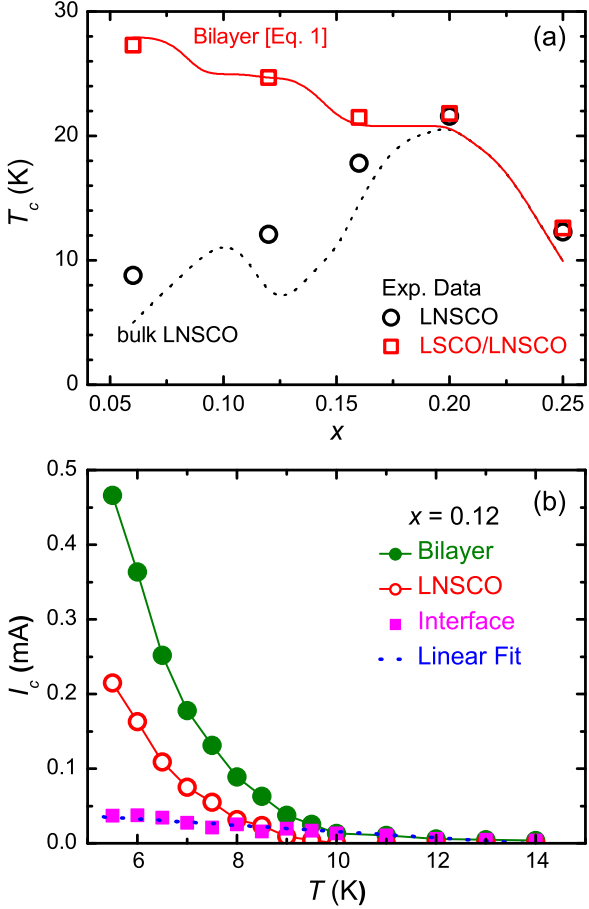}%
\end{center}
\caption{\label{fig2} (Color online) (a) The $T_c$ values as a function of Sr doping level ($x$) for LNSCO film (circle), LSCO/LNSCO bilayer (square), bulk LNSCO \cite{Crawford} (dashed line) and the prediction for bilayer using Eq. (1) (solid line). The $T_c$'s of LNSCO films are higher than that of their bulk counterparts as the films are under the compressive strain. (b) The critical current $I_{c}$($T$) for LNSCO (open circle) and bilayer films (filled circle) for $x$ = 0.12. The contribution of the interfacial layer (square), extracted from the $I_c(T)$ data, is shown with its linear fit (dotted line).}
\end{figure}

\section{Results and Discussion}

Figure 1(a-e) shows the in-plane resistivity $\rho(T)$ of LNSCO/LSCO bilayer along with the $\rho(T)$ of LNSCO monolayer films for different values of $x$. We observe distinctly higher $T_c$ for the bilayers with $x$ = 0.06, 0.12, and 0.16 as compared to that for the corresponding bare LNSCO film [See Fig. 1(a-c)]. The maximum $T_c$ enhancement of $\approx$ 18.5 K is observed for the bilayer with $x$ = 0.06. Here $T_c$ is defined as the onset temperature of the superconducting transition. For the bilayers with $x$ = 0.20 and 0.25, the LSCO capping on the LNSCO films does not affect the $T_c$ as seen in Fig. 1(d,e). Another interesting feature of our data is that the superconductivity in the bilayers can exist at a much higher temperature than in any LNSCO monolayer film [See Fig. 2(a)]. Although such kind of enhancement has been reported in LSCO based bilayers \cite{Yuli}, the $T_c$ enhancement reported here is quite large ($\sim$ 210 $\%$). The observed enhancement of $T_c$ in bilayers is not sensitive to the thickness of the individual layers or to the sequence of the layers. This suggests that the observed phenomenon is purely an interface effect. However, a natural question that emerges out is the thickness of the interface region that is being perturbed to have an enhanced $T_c$. We have estimated the thickness from the temperature dependence of critical current $I_{c}$($T$) data. Figure 2(b) shows the typical $I_{c}$($T$) curves for both monolayer and bilayer film with $x$ = 0.12. In the latter case, the $I_c$($T$) for $T \leq$ 9.0 K draws contributions from both the LNSCO layer which has a $T_c$ of $\approx$ 9.0 K and the interfacial layer with $T_c \approx$ 14.0 K, while, for the temperature between 9.0 K and 14.0 K, the $I_c$ is only because of the interface. We have assumed the bilayer to be a combination of two parallel layers of well defined thicknesses (say, $d_{LNSCO}$ and $d_{Interface}$) and $T_c$'s. The ratio between the critical currents of LNSCO and interfacial layer at $T$ = 0 K is $I_{c,LNSCO} (0)/I_{c,Interface} (0) = d_{LNSCO}/d_{Interface}$. Since $d_{LNSCO} + d_{Interface} \approx$ 100 nm for our bilayer, the thickness of the interfacial layer is $\approx$ 2.70 nm or 2 unit cells. Here we have assumed that both superconducting layers have same critical current density and are confined within the LNSCO side of bilayer. We obtain similar values (2-3 unit cells) of the interfacial layer thickness for the bilayers with $x<$ 0.20. On the other hand, the $I_c$($T$) data for both LNSCO and bilayer films are almost identical for $x=$ 0.20 and 0.25. This indicates that the superconductivity in these bilayers is due to only LNSCO layer.

\begin{figure}[htb]
\begin{center}
%\vskip -1.5cm
%\abovecaptionskip -10cm
\includegraphics [width=8cm]{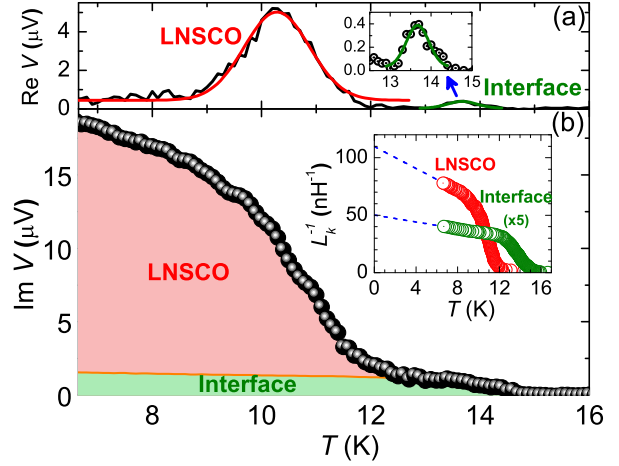}%
\end{center}
\caption{\label{fig3} (Color online) (a) The Re $V$($T$) for LNSCO($x$ = 0.12)/LSCO bilayer with the Gaussian fits showing loss peaks due to the interfacial layer (green) and LNSCO layer (red). The inset shows the magnified view of the smaller peak located at $\approx$ 13.7 K. (b) The Im $V$ vs $T$ curve for the bilayer with the contributions from the interfacial layer (green) and LNSCO layer (red). The inset shows the $L^{-1}_{k}(T)$ for the interfacial layer (multiplied by 5) and LNSCO layer. The blue dotted lines show the extrapolation of $L^{-1}_{k}(T)$ towards $T$ = 0 K.}
\end{figure}

\begin{figure*}[t]
\begin{center}
%\vskip -1.5cm
%\abovecaptionskip -10cm
\includegraphics [width=7cm]{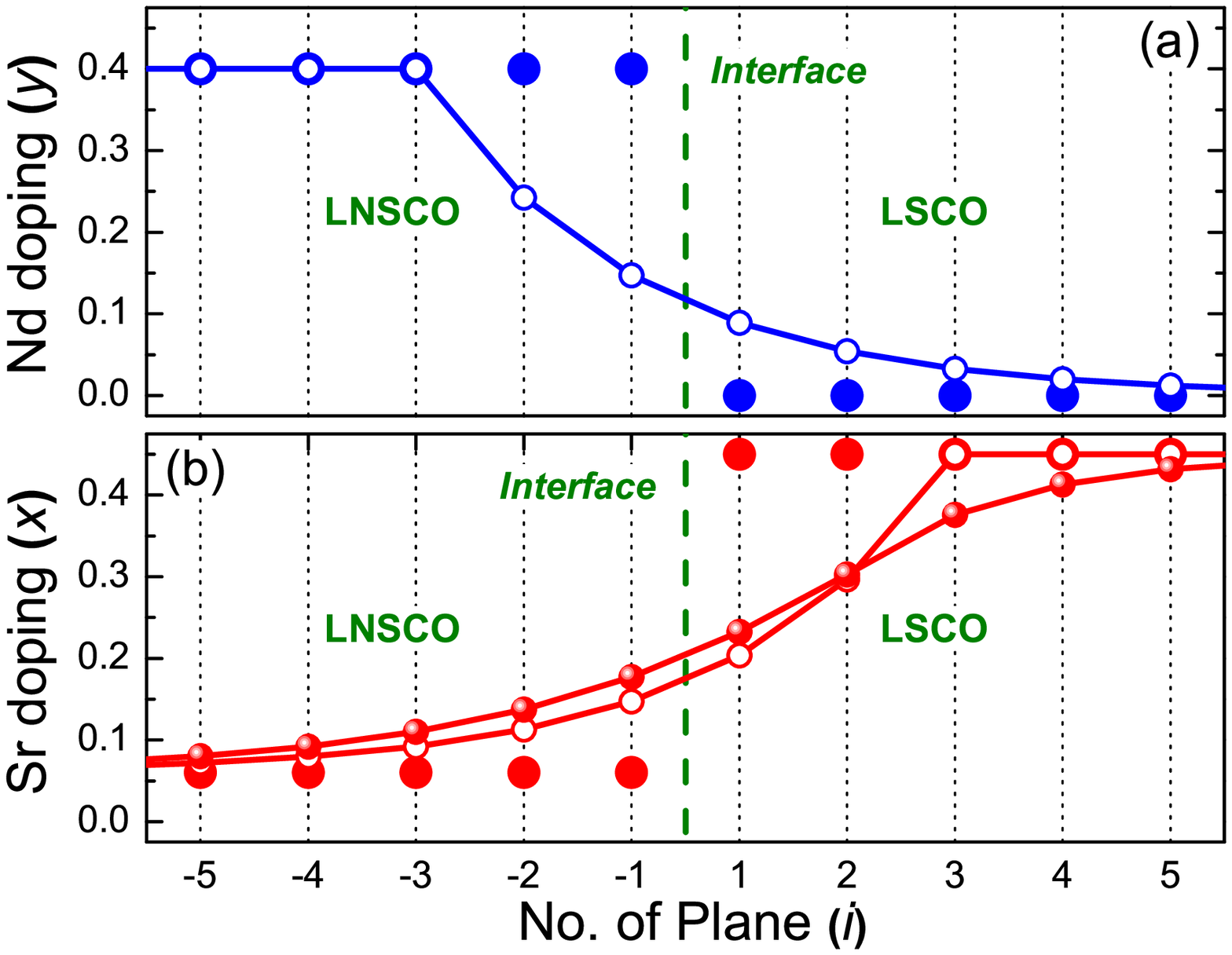}%
\hskip 0.2cm
\includegraphics [width=7.3cm]{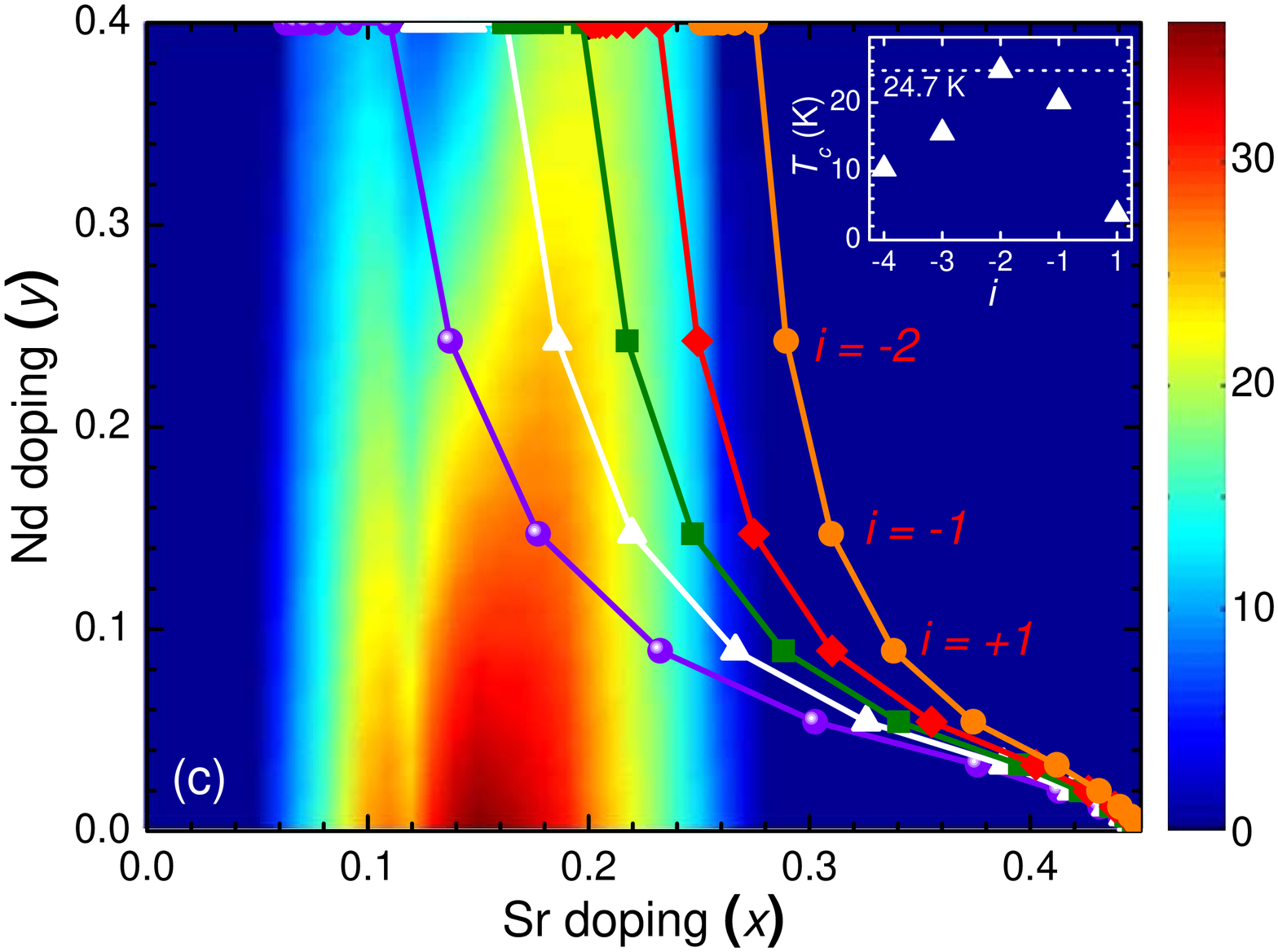}
\end{center}
\caption{\label{fig4} (Color online) (a,b) The nominal profiles (filled circle) of Sr$^{2+}$ and Nd$^{3+}$ ions across LNSCO($x$ = 0.12)/LSCO interface. The inter-diffusion of ions at the interface results in the smeared profiles (open circle). The solid spheres represent the actual hole density due to Sr$^{2+}$ doping at various CuO$_2$ planes simulated using Eq. (1). (c) The doping level ($x$ and $y$) at each CuO$_2$ planes for bilayers with $x$ = 0.06 (sphere), 0.12 (triangle), 0.16 (square), 0.20 (diamond) and 0.25 (circle). The contour map shows the superconducting onset temperature as a function of $x$ and $y$ for La$_{2-x-y}$Sr$_{x}$Nd$_{y}$CuO$_{4}$/La$_{1.55}$Sr$_{0.45}$CuO$_{4}$ bilayer systems. The values are interpolated from $T_c$ vs $x$ phase diagram of the bulk samples of La$_{1.6-x}$Nd$_{0.4}$Sr$_{x}$CuO$_{4}$ \cite{Crawford} and La$_{2-x}$Sr$_{x}$CuO$_{4}$ \cite{Takagi} assuming a linear variation of $T_c$ with $y$ \cite{Buchner}. The inset shows the $T_c$ values in the successive CuO$_2$ planes near the interface for $x$ = 0.12.}
\end{figure*}

A clear and direct evidence for the enhancement of $T_{c}$ is given by the ac screening measurements of the bilayer and Nd-doped films. The screening voltage for the bilayer with $x$ = 0.12 is shown in Fig. 3. We observe two distinct loss peaks in the real part of the pick-up coil voltage ($V$) for the bilayer. A smaller peak is located at $\approx$ 13.7 K in addition to the prominent loss peak at 10.3 K. While latter is due to the superconducting transition of LNSCO layer, as confirmed by the presence of a loss peak occurring at the same temperature in a plain LNSCO film (not shown here), the small high temperature peak can be assigned as the superconducting transition of the interfacial layer. Similarly, the deconvolution of Im $V$ data [Fig. 3(b)] of the bilayer shows two contributions corresponding to the LNSCO layer and the interfacial layer. We want to point out the fact that the enhancement of $T_c$ determined using the screening or $I_c$($T$) data is small compared to the enhancement of the onset temperature of the resistive transition. We have extracted the inverse sheet kinetic inductance $L^{-1}_{k}(T)$ of the superconducting condensate for each component from the deconvoluted data \cite{Rout} and then have calculated the Kosterlitz-Thouless-Berezinskii transition temperature ($T_{KTB}$) of the quasi two-dimensional (2D) interface layer. The $T_{KTB}$, given by the expression $L_k^{ - 1} (T_{KTB} ) = (8\pi k_B / \Phi _0^2)T_{KTB}$, where $\Phi _0$ is the flux quantum, lies close to the $T_c$. This observation further suggests the 2D nature of the interface superconducting layer. In a 2D superconductor, $L^{-1}_{k} \propto n_{s}d$, where $n_{s}$ is the superfluid density and $d$ is the layer thickness. Thus $L^{-1}_{k,LNSCO}(0)$/$L^{-1}_{k,Interface}(0) = d_{LNSCO}/d_{Interface}$ under the assumption of a uniform superfluid density in the bilayer. A rough estimate of $d_{Interface}$ can be obtained considering the two superconducting component model for the bilayer, as discussed before. Using the extrapolated values of $L^{-1}_{k}$ at 0 K for both the LNSCO and the interface layer [see the inset of Fig. 3(b)], we have extracted $d_{Interface}$ as $\approx$ 4.34 nm or 3 unit cells. Furthermore, the integrated area under the loss peak in Re $V$($T$) gives a measure of superconducting volume fraction. Thus one can get a similar value of $d_{Interface} \approx$ 3.68 nm by comparing both loss peaks of the bilayer. Although the screening measurements give an impression of a thicker interface layer as compared to that obtained from the $I_c(T)$ measurements, one must take into account the masking of the ac response of the bottom LNSCO layer by the upper superconducting interfacial layer with a higher $T_c$, which results in overestimation of $d_{Interface}$. Following all these results, it clearly appears that the observed enhanced SC is due to a few ($\sim$ 2-3) unit cells thick interfacial layer.

We have addressed the possible growth related reasons for the enhanced SC in bilayers. The close matching in the lattice parameters ($a_{LNSCO} \approx a_{LNSCO} \approx$ 0.376 nm) of two components of bilayers does not introduce any significant change in lattice structure (and the local strains) at the interface. Thus we can rule out the possibility of the strain enhancing the $T_c$ at the interface. We do not expect any excess oxygen doping, which would have substantially increased the $T_c$. The interface roughness introduces a doping inhomogeneity which can create patches of OD and UD regions within the CuO$_2$ planes at the interface. These regions can couple strongly through in-plane tunneling and thus giving rise to enhanced $T_c$ layers \cite{Goren}. But our roughness value of one unit cell limits such enhancement and thus certainly can not be a reason for the observed $T_c$ enhancement extending to few unit cells.

A simple yet satisfactory explanation of enhancement in $T_c$ can be provided by taking into account the charge transfer as well as the diffusion of the ions at the interface. Firstly, we consider the latter phenomenon. The individual entities of the bilayers contain two types of dopants, viz. Sr$^{2+}$ and Nd$^{3+}$. The concentration of these ions on both sides of the interface is different and this inequality of the concentration can trigger the physical motion of the ions from higher to lower density side. Such effect ensures a smeared ion profile instead of a sharp one \cite{Gozar,Logenov}. The ion profiles reported in Ref. [2,3] follow an empirical relation for the ion density at $j$ th plane; $n_j = n_2 + \Delta n e^{-j/\lambda}$, where $\Delta n = n_1 - n_2$ with $n_1$ and $n_2$ as the densities on both sides of the interface ($n_1 > n_2$) and $\lambda$ is the ion diffusion length. Here $j \geq$ 0 and $j=$ 0 represents the starting point of decaying ion profile from a level of $n_1$. The fits yield the value of $\lambda$ approximately equal to the interface roughness observed in these heterostructures. In our study, we have assumed that $\lambda =$ 1 unit cell, which is the roughness value for our case. Using above mentioned relation, we have simulated the ion profile due to inter-diffusion from the nominal ion densities. Figure 4(a-b) show the diffused ion profile for Nd$^{3+}$ and Sr$^{2+}$ in LNSCO($x=$ 0.12)/LSCO bilayer. Next we need to examine the charge transfer occurring at the interface. The replacement of a La$^{+3}$ by Sr$^{+2}$ will introduce a hole in CuO$_2$ plane while the Nd$^{+3}$ doping does not create any extra charge. Thus there will be a varying charge (hole) profile created due to inhomogeneous Sr$^{+2}$ doping across the interface [shown by open circles in Fig. 4(b)]. As a result, a non-uniform electric potential landscape is formed across the interface, which leads to a redistribution of electronic charges in different oxide planes. This has been modeled using Poisson's equation in discreet form \cite{Loktev}. According to this model, the modified hole density $p_{i}$ at any CuO$_2$ plane (say, {\it i} th) with initial hole density $x_{i}$ is given by the relation \cite{Loktev}:
\begin{equation}
\label{eq.1}
p_{i + 1}  + p_{i - 1}  - (2 + \alpha )p_{i}  =  - \alpha x_{i}
\end{equation}
where $\alpha$ represents the degree of charge localization in the system.  We have performed the numerical simulation of above recurrence relation to obtain the actual hole doping profile at different CuO$_2$ planes near the bilayer interface as shown in Fig 4(b) \cite{Note1}.

Figure 4(c) shows the actual doping levels in various bilayers after taking in to account cation intermixing and charge redistribution near the interface as mentioned above. Clearly we can see the oxide planes with enhanced $T_c$ for the bilayers with $x$ = 0.06, 0.12 and 0.16 while the $T_c$ remains unchanged for other bilayers. The $T_c$ profile for LNSCO($x$ = 0.12)/LSCO bilayer near the interface [inset of Fig. 4(c)] shows a $T_c$ of 24.7 K in the second LNSCO plane ($i$ = $-$2) from the interface, which is close to the observed $T_c$ in the bilayer. Thus we may conclude that the enhanced SC is due to the second CuO$_2$ plane from the LNSCO/LSCO interface on the LNSCO side. Similarly the $i$ = $-$2 layer is responsible for the enhanced $T_c$ in the bilayer with $x$ = 0.20 while $i$ = $-$1 layer for $x$ = 0.06. In case of $x$ = 0.20 and 0.25, the SC is from the bulk part of LNSCO layer ($i$ values near $-$77). The calculated values of $T_c$ matches quite well with the experimental values as shown in Fig. 2(a).

We have gone ahead to generate a generic map of $T_c$ as a function of Sr doping ($x$) and Nd doping ($y$) for La$_{2-x-y}$Sr$_{x}$Nd$_{y}$CuO$_{4}$/La$_{1.55}$Sr$_{0.45}$CuO$_{4}$ bilayer systems as shown in Fig. 5. Only for a small portion of doping values with $x <$ 0.005 and $y>$ 0.37 [top left corner of Fig. 5], a OD layer, namely $i$ = $+$1, contributes to enhanced SC. As we move away from this region, the enhanced SC layer goes deeper into LNSCO layer. After a certain level, the enhanced SC is lost and only bulk SC of LNSCO layer remains. Interestingly the CuO$_2$ plane with highest $T_c$ lies on the UD side for the majority of the phase space contrary to the other major explanation for $T_c$ enhancement, viz. the phase fluctuation model, where the enhanced SC should lie on OD side \cite{Berg}.

\begin{figure}[t]
\begin{center}
%\vskip -1.5cm
%\abovecaptionskip -10cm
\includegraphics [width=8cm]{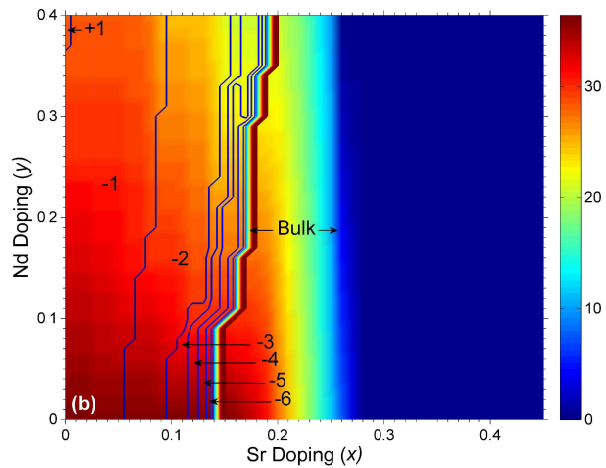}
\end{center}
\caption{\label{fig5} (Color online) The contour map showing the $T_c$ of the bilayers as predicted by Eq. (1). The phase diagram of the bilayer consists of two regions: (1) the enhanced SC region where the numbers indicate the plane number ($i$) responsible for enhanced $T_c$. and (2) the region where no enhancement is observed and the SC is only due to the bulk part of LNSCO layer (marked as ``Bulk'').}
\end{figure}

Although the above mentioned model satisfactorily explains the $T_c$ enhancement seen in our bilayers, it also predicts that  the enhanced SC is only confined within one CuO$_2$ plane (or at most two). On the contrary, we observe an anomalously thick (2-3 unit cells) interfacial layer.  We can notice that, near the layer with maximum $T_c$, i.e. $i$ = $-$2, there are few other $T_c$ values of 20.2 K and 15.6 K for $i$ = $-$1 and $-$3 planes, respectively, as shown in the inset of Fig. 3(c). Despite these values are less than eventual $T_c$ of the bilayer (or $i$ = $-$2 layer), they are still larger than that of bare LNSCO film. One would expect to see the presence of these superconducting planes as additional contributions in screening voltages. But instead we only observe the effect of bulk LNSCO layer and interfacial layer. We speculate that the $T_c$'s of few layers close to $i$ = $-$2 plane are also enhanced and thus an enhanced SC layer of few unit cells is formed. A potential explanation of this phenomenon involves the stripe fluctuations in LNSCO layer. As shown in Fig. 1, the lightly doped LNSCO films show an upturn in resistance at lower temperatures, which follows a log(1/$T$) dependence. The logarithmic insulating nature has been observed due to stripe ordering in La$_{2-x-y}$Nd$_y$Sr$_{x}$CuO$_{4}$ \cite{Noda}. With increasing the doping level, the stripe fluctuation increases and ultimately two-dimensional metal is formed at overdoped side \cite{Liu,Ando}. The presence of dynamically fluctuating stripes in LSCO can weaken the stripe order in few layers of LNSCO near the interface, if not more, and thus promoting enhancement in $T_c$. Another possibility is the proximity effect due to which the SC can be enhanced in few more interfacial layers. Gariglio \textit{et. al.} have suspected the existence of the proximity effect in La$_{1.56}$Sr$_{0.44}$CuO$_{4}$/La$_{2}$CuO$_{4}$ bilayer \cite{Gariglio}. The superconductivity in the bilayer lies only at the second CuO$_2$ plane in La$_{2}$CuO$_{4}$ from the interface \cite{Logenov}. The Zn doping in this plane suppressed the $T_c$ considerably as expected. Surprisingly, the doping in few neighboring non-superconducting planes reduced the $T_c$ albeit by smaller amount. This suggests some kind of coupling between the planes present near the superconducting plane.

\section{Conclusion}

In summary, the bilayers consisting of superconducting La$_{1.6-x}$Nd$_{0.4}$Sr$_{x}$CuO$_{4}$ capped by the overdoped metallic La$_{1.55}$Sr$_{0.45}$CuO$_{4}$ have a superconducting transition at a higher temperature as compared to the $T_c$ of bare LNSCO films for $x<$ 0.20. The ac screening measurements show two distinct superconducting transitions in such bilayers with the higher $T_c$ phase corresponding to the interfacial layer. The maximum $T_c$ enhancement of $\approx$ 18.5 K (or 210 $\%$) is seen for the bilayer with $x$ = 0.06. Contrary the $T_c$ of the bilayers with $x\geq$ 0.20 remains unchanged. The enhancement of superconductivity is confined to $\approx$ 2-3 unit cells thick interfacial layer, as evidenced by the temperature dependent critical current and ac screening data. The enhancement observed in these bilayers can be explained by a model considering the charge redistribution due to the potential differences and the cation inter-diffusion across the interface. Still the reason of unusually large interfacial layer thickness is unclear. The relaxation of stripe order in La$_{1.84}$Sr$_{0.16}$CuO$_{4}$ and the proximity effect are crucial factors for further understanding of the interface superconductivity in cuprate heterostructures.

\acknowledgments
P.K.R. acknowledges the financial support from the Council of Scientific and Industrial Research, Government of India. R.C.B. acknowledges the J. C. Bose fellowship of Department of Science and Technology, India.


\begin{thebibliography}{0}

\bibitem{Zubko}
  \Name{Zubko P., Gariglio S., Gabay M., Ghosez P. \and Triscone J.-M.}
  \REVIEW{Annu. Rev. Condens. Matter Phys.}{2}{2011}{141}.

\bibitem{Gozar}
  \Name{Gozar A., Logvenov G., Kourkoutis L. F., Bollinger A. T., Giannuzzi L. A., Muller D. A. \and Bozovic I.}
  \REVIEW{Nature}{55}{2008}{782}.

\bibitem{Logenov}
  \Name{Logvenov G., Gozar A. \and Bozovic I.}
  \REVIEW{Science}{326}{2009}{699}.

\bibitem{Smadici}
  \Name{Smadici S., Lee J. C. T., Wang S., Abbamonte P., Logvenov G., Gozar A., Cavellin C. D. \and Bozovic I.}
  \REVIEW{SPhys. Rev. Lett.}{102}{2009}{107004}.

\bibitem{Yuli}
  \Name{Yuli O., Asulin  I., Millo O., Orgad D., Iomin L. \and Koren G.}
  \REVIEW{Phys. Rev. Lett.}{101}{2008}{057005}.
  \Name{Koren G. \and Millo O.}
  \REVIEW{Phys. Rev. B}{81}{2010}{134516}.

\bibitem{Rout}
  \Name{Rout P. K. \and Budhani R. C.}
  \REVIEW{Phys. Rev. B}{82}{2010}{024518}.

\bibitem{Jin}
  \Name{Jin K., Bach P., Zhang X. H., Grupel U., Zohar E., Diamant I., Dagan Y., Smadici S., Abbamonte P. \and Greene R. L.}
  \REVIEW{Phys. Rev. B}{83}{2011}{060511(R)}.

\bibitem{Locquet}
  \Name{Locquet J.-P., Perret J., Fompeyrine J., Mächler E., Seo J. W. \and Tendeloo G. V.}
  \REVIEW{Nature}{394}{1998}{453}.

\bibitem{Bozovic}
  \Name{Bozovic I., Logvenov G., Belca I., Narimbetov B. \and Sveklo I.}
  \REVIEW{Phys. Rev. Lett.}{89}{2002}{107001}.
  \Name{Mohottala H. E., Wells B. O., Budnick J. I., Hines W. A., Niedermayer C., Udby L., Bernhard C., Moodenbaugh A. R. \and Chou F. C.}
  \REVIEW{Nature Mater.}{5}{2006}{377}.

\bibitem{Berg}
  \Name{Berg E., Orgad D. \and Kivelson  S. A.}
  \REVIEW{Phys. Rev. B}{78}{2008}{094509}.
  \Name{Okamoto S. \and Maier T. A.}
  \REVIEW{Phys. Rev. Lett.}{101}{2008}{156401}.
  \Name{Goren L. \and Altman E.}
  \REVIEW{Phys. Rev. B}{79}{2009}{174509}.

\bibitem{Loktev}
  \Name{Loktev V. M. \and Pogorelov Y. G.}
  \REVIEW{Phys. Rev. B}{78}{2008}{180501(R)}.

\bibitem{Tranquada}
  \Name{Tranquada J. M., Sternlieb B. J., Axe J. D., Nakamura Y. \and Uchida S.}
  \REVIEW{Nature}{375}{1995}{561}.

\bibitem{Liu}
  \Name{Liu Y., Qu J. F., Zhu M., Zhang S. D., Feng S. J. \and Li X. G.}
  \REVIEW{Phys. Rev. B}{70}{2004}{224512}.

\bibitem{Tsukada}
  \Name{Tsukada I.}
  \REVIEW{Phys. Rev. B}{64}{2001}{224501}.

\bibitem{Wakimoto}
  \Name{Wakimoto S., Zhang H., Yamada K., Swainson I., Kim H. \and Birgeneau R. J.}
  \REVIEW{Phys. Rev. Lett.}{92}{2004}{217004}.

\bibitem{Kivelson}
  \Name{Kivelson S. A., Fradkin E. \and Emery V. J.}
  \REVIEW{Nature}{393}{1998}{550}.

\bibitem{Ando}
  \Name{Ando Y., Lavrov A. N., Komiya S., Segawa K. \and Sun X. F.}
  \REVIEW{Phys. Rev. Lett.}{87}{2001}{017001}.
  
\bibitem{Dane}
  \Name{Dane A. D., Veldhuis A., de Boer D. K. G., Leenaers A. J. G. \and Buydens L. M. C.}
  \REVIEW{Physica B}{253}{1998}{254}.

\bibitem{Jeanneret}
  \Name{Jeanneret B., Gavilano J. L., Racine G. A., Leemann Ch. \and Martinoli P.}
  \REVIEW{Appl. Phys. Lett.}{55}{1989}{2336}.

\bibitem{Crawford}
  \Name{Crawford M. K., Harlow R. L., McCarron E. M., Farneth W. E., Axe J. D., Chou H. \and Huang Q.}
  \REVIEW{Phys. Rev. B}{44}{1991}{7749}.

\bibitem{Goren}
  \Name{Goren L. \and Altman E.}
  \REVIEW{Phys. Rev. B}{84}{2011}{094508}.

\bibitem{Note1}
The numerical simulations are done with the value of $i$ ranging from $-$38 to 38, corresponding to 50 nm of LNSCO and 50 nm of LSCO layer thickness so that $i<$ 0 represents the CuO$_2$ planes in LNSCO side while $i>$ 0, the planes in LSCO side. We have used an empirical value of $\alpha$ = 0.5. A finite stack of CuO$_2$ layers in bilayers introduces the boundary conditions: $x_i$ = 0 = $p_i$ for $i$ = $-$39 (substrate side) and 39 (free space side).

\bibitem{Takagi}
  \Name{Takagi H., Ido T., Ishibashi S., Uota M., Uchida S. \and Tokura Y.}
  \REVIEW{Phys. Rev. B}{40}{1989}{2254}.

\bibitem{Buchner}
  \Name{B\"{u}chner B., Breuer M., Freimuth A. \and Kampf A. P.}
  \REVIEW{Phys. Rev. Lett.}{73}{1994}{1841}.

\bibitem{Noda}
  \Name{Noda T., Eisaki H. \and Uchida S.}
  \REVIEW{Science}{286}{1999}{265}.

\bibitem{Gariglio}
  \Name{Gariglio S., Gabay M. \and Triscone J.-M.}
  \REVIEW{Nat. Nanotechnol.}{5}{2010}{13}.


\end{thebibliography}
\end{document}